\documentclass[runningheads]{llncs}
\pdfoutput=1
\usepackage{tabularx}
\usepackage{graphicx}
\usepackage{adjustbox}
\usepackage{algorithm}
\usepackage{algpseudocode}
\usepackage{amsmath}
\usepackage{multirow}
\usepackage{makecell}
\usepackage[super]{nth}

\usepackage{todonotes}
\usepackage{pgfplots}
\usepackage{pgfplotstable}
\usepgfplotslibrary{groupplots}

\pgfplotsset{compat=1.7}
\usepackage{tikz}
\usepackage[square, numbers]{natbib}
\usepackage{comment}
\usepackage{soul}

\usepackage{caption}
\usepackage{subcaption}

\makeatletter
\newcommand\avsuminner[2]{%
  {\sbox0{$\m@th#1\sum$}%
   \vphantom{\usebox0}%
   \ooalign{%
     \hidewidth
     \smash{\vrule height\dimexpr\ht0+1pt\relax depth\dimexpr\dp0+1pt\relax}%
     \hidewidth\cr
     $\m@th#1\sum$\cr
   }%
  }%
}
\makeatother

\newcolumntype{P}[1]{>{\centering\arraybackslash}p{#1}}
\newcolumntype{M}[1]{>{\centering\arraybackslash}m{#1}}

\begin{document}

\title{New Metrics to Encourage Innovation and Diversity in Information Retrieval Approaches} 
\titlerunning{New Metrics to Encourage Innovation and Diversity in IR} 

\author{Mehmet Deniz Türkmen\inst{1} \and
Matthew Lease\inst{2,} \and
Mucahid Kutlu\inst{1}}

\institute{Dept. of Computer Eng., TOBB U.\ of Economics \& Technology, Ankara, Turkey  \and School of Information, University of Texas at Austin, USA  \\
m.turkmen@etu.edu.tr, m.kutlu@etu.edu.tr, ml@utexas.edu}

\authorrunning{M.D.\ Türkmen, M.\ Lease, and M.\ Kutlu}

\maketitle

\begin{abstract}

In evaluation campaigns, participants often explore variations of popular, state-of-the-art baselines as a low-risk strategy to achieve competitive results. While effective, this can lead to local ``hill climbing’’ rather than a more radical and innovative departure from standard methods. Moreover, if many participants build on similar baselines, the overall diversity of approaches considered may be limited. In this work, we propose a new class of IR evaluation metrics intended to promote greater diversity of approaches in evaluation campaigns. Whereas traditional IR metrics focus on user experience, our two ``innovation'' metrics instead reward exploration of more divergent, higher-risk strategies finding relevant documents missed by other systems. Experiments on four TREC collections show that our metrics do change system rankings by rewarding systems that find such rare, relevant documents. This result is further supported by a controlled, synthetic data experiment, and a qualitative analysis. In addition, we show that our metrics achieve higher evaluation stability and discriminative power than the standard metrics we modify. To support reproducibility, we share our source code.

\keywords{Evaluation, Metrics, Information Retrieval}
\end{abstract}

\section{Introduction}
\vspace{-1em}

Researchers must balance risk vs. reward in prioritizing methods to investigate. Higher-risk methods offer  the potential for a larger impact, but with a greater chance of sub-baseline performance. In contrast, lower-risk methods are more likely to yield improvement but may be incremental. A popular strategy to straddle such risk is to investigate variants of popular state-of-the-art models (e.g., use of pre-trained language models, such as GPT-3 \cite{brown2020language}). While this represents a low-risk strategy to achieve competitive results, it can lead to local ``hill climbing’’ rather than exploring higher-risk, more radical departures from current state-of-the-art methods. Moreover, if many researchers build on similar baselines, this can limit the overall diversity of approaches being explored in the field.

In this work, we investigate a novel class of ``innovation'’ evaluation metrics that seek to promote greater diversity among participant methods in evaluation campaigns. Such community benchmarking and evaluation campaigns play an important role in assessing the current state-of-the-art and promoting continuing advancements. For participants, evaluation campaigns provide a valuable testing ground for novel methods, and evaluation metrics chosen by a campaign can galvanize community attention on particular aspects of system performance. Evaluation campaign metrics thus help to steer a field.

Whereas traditional IR metrics focus on ranking quality for the user, our innovation metrics instead reward exploration of more divergent, higher-risk ranking methods that find relevant documents missed by most other systems. The key intuition is that a system finding relevant documents missed by other systems must differ in approach. Specifically, we modify standard Precision@K and Average Precision metrics to reward retrieval of such ``rare’’ relevant documents missed by other systems. A simple mixture-weight parameter controls the relative weight placed on such rarity, and setting this to zero reverts to the original metric. As such, evaluation campaigns adopting our metrics could easily control the extent to which they want to reward diversity of approaches vs.\ more standard user-oriented performance measures. 

Experiments over four TREC collections show that our proposed metrics do yield different rankings of systems compared to the existing metrics. In particular, we observe a steady decrease in rank correlation with official system rankings as greater weight is placed on finding rare, relevant documents. This means that if our metrics were adopted in practice, participants would be incentivized to retrieve more diverse relevant documents, with the potential to spur further innovation in the field. Additional results show that our metrics provide higher discriminative power and evaluation stability than the standard Precision@K and Average Precision metrics that we modify. 

{\bf Contributions}. 1) We propose a novel class of ``innovation’’ metrics to stimulate greater diversity of document ranking approaches for evaluation campaigns. Future work is expected to expand and improve upon our initial metrics. 2) We propose new generalizations of classic P@K and AP metrics via a simple user-specified mixture weight. This allows weighting document rarity or trivial reversion to the standard metric. 3) Results over four TREC collections show our metrics change system rankings, as well as providing higher discriminative power and evaluation stability than the standard metrics we modify. 4) We share our source code to support reproducibility and follow-on work\footnote{ \url{https://github.com/mdenizturkmen/ecir2023}}.

Our article is organized as follows. Section \ref{sec_proposed} describes our proposed metrics. Section \ref{sec_exp_syn} then presents an initial, controlled study using synthetic data to show how retrieving rare vs.\ common documents affects system rankings. Next, Section \ref{sec_exp_real} presents our main results with TREC collections, including a qualitative analysis in Section \ref{sec_exp_qual}. We then present discussion and limitations in Section \ref{sec_disc}. Section \ref{sec_rel} discusses related work, and we conclude in Section \ref{sec_conc}.

\section{Proposed IR Metrics}\label{sec_proposed}
\vspace{-1em}

Retrieval of {\em rare} documents (that few or no other systems retrieve) indicates that a system's ranking algorithm diverges from that of other systems. In this section, we introduce our two ``innovation'' metrics that seek to promote exploration of different approaches by rewarding retrieval of such rare, relevant documents. Specifically, we adapt 
Precision@K (P@K) (Section \ref{sec_hp}) and Average Precision (AP) metrics (Section \ref{sec_hwhp}), introducing a linear interpolation parameter $\alpha$ that balances the original metric vs.\ innovation by varying the weight placed on document rarity. In both cases, setting $\alpha=0$ reverts to the original metric.

\vspace{-0.5em}
\subsection{Rareness-Based Precision@K ($P@K_{Rareness}$)}\label{sec_hp}
\vspace{-0.5em}

We define our rareness-based precision-at-k as follows:

\vspace{-1em}
\begin{equation}
     P@K_{Rareness} = \frac{1}{k}\sum_{i=1}^{k} Rel(d_i) \, (1 + \alpha \, R(d_i))
\end{equation}

\noindent
where $k$ is the rank cut-off value, $Rel(d_i)$ is a binary indicator function for whether $d_i$ is relevant or not, $R(d_i)$ quantifies document rarity, and $\alpha$ is the aforementioned linear interpolation parameter. As noted earlier, setting $\alpha=0$ reverts to the standard P@K formula. In the other direction, larger $\alpha$ values provide greater rewards for  the retrieval of rare documents. Like the original P@K, only relevant documents contribute to the score (i.e., when $Rel(d_i)=1$), so document rarity is immaterial when $Rel(d_i)=0$.
We define rarity $R(d)$ by:
\vspace{-0.5em}
\begin{equation}\label{eq_2}
    R(d) = 1 - \frac{S_d}{S}
\vspace{-0.5em}
\end{equation}
where $S$ is the total number of systems and $S_d \geq 1$ is the number of those that retrieve document $d$. Rareness is bounded by $R(d) \in [0, \frac{(S-1)}{S}]$, minimized when a document is retrieved by all systems (i.e., $S_d = S$) and maximized when only one system retrieves $d$ (i.e., $S_d=1$). Therefore, as the number of systems $S$ increases, 
retrieving rare documents becomes more valuable. 

While $\alpha$ can be at any value, we recommend setting $\alpha \in [0,1]$ yielding bounds of $P@K_{Rareness} \in [0,2)$. The lower bound of $P@K_{Rareness} = 0$ occurs when all documents are non-relevant. 
The upper-bound is reached when $\alpha=1$ and all retrieved documents are relevant and have maximal rarity $R(d) = \frac{(S-1)}{S}$, thus $P@K_{Rareness} = 2\frac{(S-1)}{S} < 2$.


\vspace{-0.5em}
\subsection{Rareness Based Average Precision ($AP_{Rareness}$)}\label{sec_hwhp}
\vspace{-0.5em}

Assuming  $N_R$ relevant documents for a given topic, we define $AP_{Rareness}$ as:

\vspace{-0.5em}
\begin{equation}\label{eq_3}
    AP_{Rareness} = \frac{1}{N_R} \sum_{i=1}^{k} Rel(d_i) P@K_{Rareness}(i)
\end{equation}

When $\alpha=0$, $P@K_{Rareness} = P@K$, and thus $AP_{Rareness} = AP$. $AP_{Rareness}$ directly inherits $P@K_{Rareness}$'s same lower-bound and upper-bound of $[0,2)$.





\vspace{-1em}
\section{Experiment with Synthetic data} \label{sec_exp_syn}
\vspace{-1em}

We first present a controlled, synthetic data experiment to explore the behavior of $P@K_{Rareness}$ for varying $\alpha \in [0,1]$ and numbers of $D$ relevant documents retrieved. We contrast the evaluation of two hypothetical systems: $S_{rare}$ vs.\ $S_{common}$, on a single topic (\#1127540) from the Deep Learning Track 2020 (DLT20) \cite{craswell2021overview}, as if our hypothetical systems had participated with other real participants. While $S_{rare}$ always retrieves simulated relevant documents found by no other system, $S_{common}$ retrieves the most common, real relevant documents first.  
We include all official runs from DLT20's document ranking task.

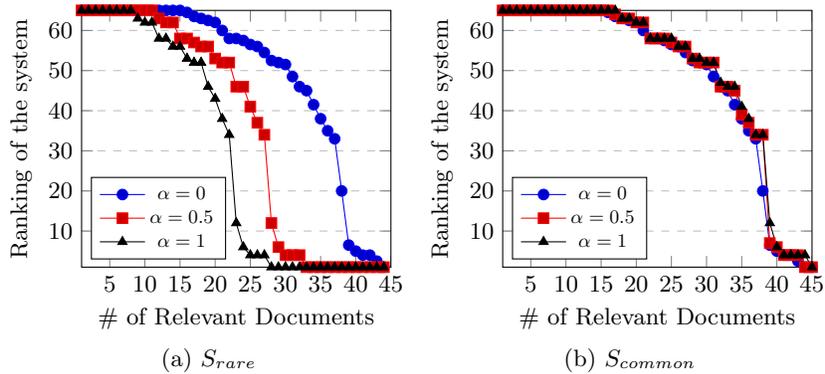
\begin{figure}[!t]
\centering
\begin{subfigure}[b]{0.45\textwidth}
         \begin{tikzpicture}
\begin{axis}[
    scale= 0.6,
    xlabel={\# of Relevant Documents},
    ylabel={Ranking of the system},
    xmin=1, xmax=45,
    ymin=1, ymax=65,
    xtick={0,5,10,15,20,25,30,35,40,45},
    ytick={0,10,20,30,40,50,60,70},
    legend pos=south west,
    ymajorgrids=true,
    grid style=dashed,
    legend style={nodes={scale=0.8, transform shape}}, 
]\addplot
    coordinates {
    (1,65.0)(2,65.0)(3,65.0)(4,65.0)(5,65.0)(6,65.0)(7,65.0)(8,65.0)(9,65.0)(10,65.0)(11,65.0)(12,65.0)(13,65.0)(14,65.0)(15,65.0)(16,64.5)(17,63.5)(18,63.0)(19,62.5)(20,62.0)(21,60.0)(22,58.0)(23,58.0)(24,57.5)(25,56.5)(26,56.0)(27,54.5)(28,52.5)(29,52.0)(30,51.5)(31,48.5)(32,46.0)(33,45.0)(34,41.5)(35,38.0)(36,35.0)(37,33.0)(38,20.0)(39,6.5)(40,5.0)(41,4.0)(42,4.0)(43,2.5)(44,1.0)
    };
    
    
    
 
\addplot  
    coordinates {
    (1,65.0)(2,65.0)(3,65.0)(4,65.0)(5,65.0)(6,65.0)(7,65.0)(8,65.0)(9,65.0)(10,65.0)(11,65.0)(12,63.0)(13,62.0)(14,62.0)(15,58.0)(16,58.0)(17,57.0)(18,56.0)(19,56.0)(20,53.0)(21,52.0)(22,52.0)(23,46.0)(24,46.0)(25,41.0)(26,37.0)(27,34.0)(28,12.0)(29,6.0)(30,4.0)(31,4.0)(32,4.0)(33,1.0)(34,1.0)(35,1.0)(36,1.0)(37,1.0)(38,1.0)(39,1.0)(40,1.0)(41,1.0)(42,1.0)(43,1.0)(44,1.0)
    };
 
    
\addplot  +[mark options={scale=1,fill=black},mark=triangle*,black ]
    coordinates {
    (1,65.0)(2,65.0)(3,65.0)(4,65.0)(5,65.0)(6,65.0)(7,65.0)(8,65.0)(9,63.0)(10,62.0)(11,62.0)(12,58.0)(13,58.0)(14,56.0)(15,56.0)(16,53.0)(17,52.0)(18,52.0)(19,46.0)(20,43.0)(21,38.0)(22,34.0)(23,12.0)(24,6.0)(25,4.0)(26,4.0)(27,4.0)(28,1.0)(29,1.0)(30,1.0)(31,1.0)(32,1.0)(33,1.0)(34,1.0)(35,1.0)(36,1.0)(37,1.0)(38,1.0)(39,1.0)(40,1.0)(41,1.0)(42,1.0)(43,1.0)(44,1.0)
    };
    \legend{$\alpha=0$,$\alpha=0.5$,$\alpha=1$,}
\end{axis}
\end{tikzpicture}
         \vspace{-1.5em}
         \caption{$S_{rare}$}
     \end{subfigure}
\begin{subfigure}[b]{0.45\textwidth}
         \begin{tikzpicture}
\begin{axis}[
    scale= 0.6,
    xlabel={\# of Relevant  Documents},
    ylabel={Ranking of the system},
    xmin=1, xmax=45,
    ymin=1, ymax=65,
    xtick={0,5,10,15,20,25,30,35,40,45},
    ytick={0,10,20,30,40,50,60,70},
    legend pos=south west,
    ymajorgrids=true,
    grid style=dashed,
    legend style={nodes={scale=0.8, transform shape}}, 
]\addplot 
    coordinates {
    (1,65.0)(2,65.0)(3,65.0)(4,65.0)(5,65.0)(6,65.0)(7,65.0)(8,65.0)(9,65.0)(10,65.0)(11,65.0)(12,65.0)(13,65.0)(14,65.0)(15,65.0)(16,64.5)(17,63.5)(18,63.0)(19,62.5)(20,62.0)(21,60.0)(22,58.0)(23,58.0)(24,57.5)(25,56.5)(26,56.0)(27,54.5)(28,52.5)(29,52.0)(30,51.5)(31,48.5)(32,46.0)(33,45.0)(34,41.5)(35,38.0)(36,35.0)(37,33.0)(38,20.0)(39,6.5)(40,5.0)(41,4.0)(42,4.0)(43,2.5)(44,1.0)(45,1.0)

    };
    
    
    
    
\addplot 
    coordinates {
    (1,65.0)(2,65.0)(3,65.0)(4,65.0)(5,65.0)(6,65.0)(7,65.0)(8,65.0)(9,65.0)(10,65.0)(11,65.0)(12,65.0)(13,65.0)(14,65.0)(15,65.0)(16,65.0)(17,64.0)(18,63.0)(19,63.0)(20,62.0)(21,62.0)(22,58.0)(23,58.0)(24,58.0)(25,57.0)(26,56.0)(27,56.0)(28,53.0)(29,52.0)(30,52.0)(31,52.0)(32,46.0)(33,46.0)(34,45.0)(35,39.0)(36,37.0)(37,34.0)(38,34.0)(39,7.0)(40,6.0)(41,4.0)(42,4.0)(43,4.0)(44,1.0)(45,1.0)
    };
    
\addplot +[mark options={scale=1,fill=black},mark=triangle*,black ]
    coordinates {
    (1,65.0)(2,65.0)(3,65.0)(4,65.0)(5,65.0)(6,65.0)(7,65.0)(8,65.0)(9,65.0)(10,65.0)(11,65.0)(12,65.0)(13,65.0)(14,65.0)(15,65.0)(16,65.0)(17,65.0)(18,63.0)(19,63.0)(20,62.0)(21,62.0)(22,58.0)(23,58.0)(24,58.0)(25,58.0)(26,56.0)(27,56.0)(28,53.0)(29,53.0)(30,52.0)(31,52.0)(32,47.0)(33,46.0)(34,46.0)(35,41.0)(36,38.0)(37,34.0)(38,34.0)(39,12.0)(40,6.0)(41,4.0)(42,4.0)(43,4.0)(44,4.0)(45,1.0)
    };
    \legend{$\alpha=0$,$\alpha=0.5$,$\alpha=1$,}
\end{axis}
\end{tikzpicture}
         \vspace{-1.5em}
         \caption{$S_{common}$}
     \end{subfigure}
\vspace{-1em}
\caption{Ranking of hypothetical systems, $S_{rare}$ and $S_{common}$, for topic 1127540 of Deep Learning Track 2020 based on $P@100_{Rareness}$. Experiments vary rarity weight $\alpha$ as well as $D$, the number of relevant documents retrieved.}
\label{fig:synthetic_exp_alpha}
\vspace{-1.5em}
\end{figure}

\textbf{Figure \ref{fig:synthetic_exp_alpha}} shows the $P@100_{Rareness}$ ranking of $S_{rare}$ vs.\ $S_{common}$. 
Note that a lower rank indicates a better system, with the best system being ranked first (i.e., having rank 1). 
First, recall that when $\alpha=0$, $P@K_{Rareness} = P@k$. In this case, both $S_{rare}$ and $S_{common}$ are seen to exhibit the same $P@K_{Rareness}$ curve, as expected, since no weight is placed on rarity. 
%
%
Second, we see that $S_{common}$'s ranking is largely unaffected by $\alpha$ since it always retrieves common (i.e., non-rare) relevant documents. 
In contrast, the ranking of $S_{rare}$ noticeably changes across different $\alpha$ values. For example, it requires 28, 33, and 44 relevant documents to be ranked first when $\alpha$ is set to 1, 0.5, and 0, respectively. 

Overall, the results above validate our expectations regarding the behavior of $P@k_{Rareness}$ under controlled conditions. It reverts toward standard $P@k$ at $\alpha=0$, 
and results place greater emphasis on rarity as we move toward $\alpha=1$.
 
\vspace{-1em}
\section{Experiments with Real Data} \label{sec_exp_real}
\vspace{-0.75em}

In this section, we first describe our experimental setup (Section \ref{sec_exp_setup}). Next, we compare our modified metrics vs.\ their original counterparts in terms of system rankings (Section \ref{sec_exp_rank}), discriminative power (Section \ref{sec_exp_disc}), and evaluation stability (Section \ref{sec_exp_rel}). We also assess how our metrics are affected by the number of systems (Section \ref{sec_exp_system}). Furthermore, we conduct qualitative analysis to better understand the nature of rarely-retrieved documents (Section \ref{sec_exp_qual}).

\vspace{-0.75em}
\subsection{Experimental Setup}\label{sec_exp_setup}
\vspace{-0.5em}

We use trec\_eval\footnote{https://trec.nist.gov/trec\_eval/} for calculations of classical evaluation metrics.  We set the cut-off threshold to 100 for all metrics we use including ours.
We use four different TREC collections, including  TREC-5 \cite{harman1996overview}, TREC-8 \cite{trec8}, Web Track 2014 (WT14) \cite{wt14}, and Deep Learning Track 2020 (DLT20) \cite{craswell2021overview}. 
We carry out our experiments using all official runs from ad-hoc search tasks of TREC-5, TREC-8, and WT14, and the document ranking task of DLT20.

\subsection{System Rankings}\label{sec_exp_rank}
\vspace{-0.5em}

We compare system rankings for $P@100_{Rareness}$ and $AP_{Rareness}$  against rankings based on P@100 and AP, respectively, in order to observe the impact of rewarding rarity.  We report Kendall's $\tau$ rank correlation. Experiments with $\tau_{AP}$ \cite{tauap} yielded similar results and so are omitted. 

\begin{figure*}[t]
\centering
\pgfplotsset{
   grid=both,
  minor tick num=1,
  xtick={0, 0.5, 1.0},
 ytick={0.8,0.9,1.0},
  enlargelimits=0.02,
   every tick label/.append style={font=\small},
  group style={
    columns=4,
    xlabels at=edge bottom,
    ylabels at=edge left},
  every axis legend/.append style={
    legend cell align=left,
  }
}

\begin{tikzpicture}
\begin{groupplot}[group style={group size= 4 by 1, horizontal sep=0.3cm},
        height              = 4.5cm, 
        width=4.1cm,
        xlabel              = $\alpha$ ,
        legend entries={AP vs. $AP_{Rareness}$, P@100 vs. $P@100_{Rareness}$ } ,
        legend style={
            at={(1.5,4)},
            anchor=north west,
            legend columns=-1,transpose legend,
            nodes={font=\small},
            /tikz/every even column/.append style={column sep=0.1cm}
        },
        grid=both,   
	    yticklabel style    = {/pgf/number format/precision=4}  ,
	    scaled y ticks      = false,
         every axis title shift=0,
      legend to name=grouplegend
        ]

\nextgroupplot[title=TREC5,
 ylabel style={align=center}, 
ylabel={Kendall's $\tau$},ylabel shift = 0 pt,ymin=0.75,ymax=1]   
 \addplot +[  mark options={scale=1.7} ] coordinates {(0,1.0)  (0.25,0.984)(0.5,0.965)(0.75,0.952)(1,0.942)};

        \addplot coordinates {(0,1.0)(0.25,0.973)(0.5,0.965)(0.75,0.954)(1,0.942)};
    \coordinate (top) at (rel axis cs:0,1.1);

\nextgroupplot[title=TREC8,yticklabels = \empty,ymin=0.75,ymax=1]   

\addplot +[  mark options={scale=1.7} ] coordinates {(0,1.0)  (0.25,0.993)(0.5,0.984)(0.75,0.976)(1,0.970)};

\addplot coordinates {(0,0.999)(0.25,0.992)(0.5,0.982)(0.75,0.976)(1,0.971)};

\nextgroupplot[title=WT14,yticklabels = \empty,ylabel shift = -7 pt,ymin=0.75,ymax=1]   

\addplot  +[  mark options={scale=1.7} ] coordinates {(0,1.0)  (0.25,0.990)(0.5,0.960)(0.75,0.926)(1,0.916)};

\addplot coordinates {(0,1.0)(0.25,0.974)(0.5,0.971)(0.75,0.974)(1,0.950)};
    \coordinate (top) at (rel axis cs:0,1.1);

\nextgroupplot[title=DLT20,yticklabels = \empty, 
ymin=0.75,ymax=1]

\addplot +[  mark options={scale=1.7} ] coordinates {(0,1.0)  (0.25,0.986)(0.5,0.969)(0.75,0.961)(1,0.943)};

\addplot coordinates {(0,0.994)(0.25,0.846)(0.5,0.812)(0.75,0.803)(1,0.773)};
    \coordinate (top) at (rel axis cs:0,1.1);

\end{groupplot}


    \ref{grouplegend}
\end{tikzpicture}
\vspace{-1em}
\caption{Kendall's $\tau$ correlation between system rankings based on $P@100_{Rareness}$ vs. $P@100$  and system rankings based on AP vs.  $AP_{Rareness}$. 
} 
\label{fig:ranking_p_rareness}
\vspace{-2em}
\end{figure*}
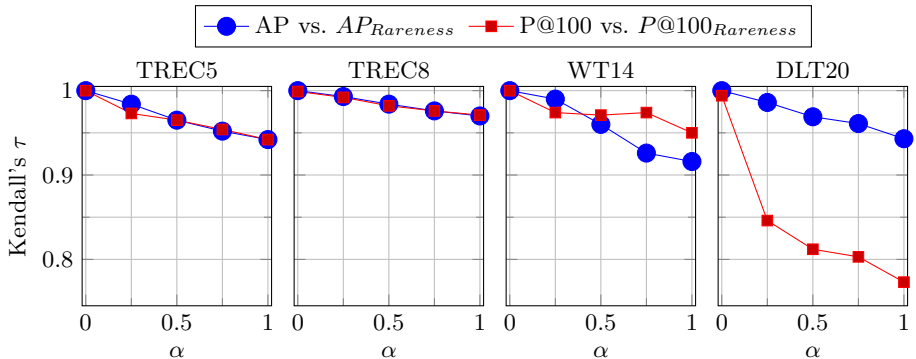

\textbf{Figure \ref{fig:ranking_p_rareness}} 
shows Kendall's $\tau$ scores  on four test collections for varying $\alpha$. 
As expected, when $\alpha$=0, our modified metrics revert to their unmodified forms, thus yielding perfect $\tau=1$ rank correlation. 
We observe steady trends of decreasing rank correlation with increasing $\alpha$.  While Kendall's  $\tau$ scores for comparisons against AP and P@K metrics are similar in TREC-5 and TREC-8, they diverge in WT14 and DLT20. For instance,  when we compare P@100 vs. $P@100_{Rareness}$ in DLT20, Kendall's $\tau$ is lower than 0.9 (a traditionally-accepted threshold for acceptable correlation \cite{voorhees2000variations})  for $\alpha > 0$. However, we do not observe this when we compare $AP_{Rareness}$  vs. AP. This suggests that DLT20 systems retrieve many rare, relevant documents at low ranks, causing large changes in system rankings when we use $P@100_{Rareness}$. Smaller changes occur with $AP_{Rareness}$ as the impact of documents is diminished due to their low ranks. 


\subsection{Discriminative Power}\label{sec_exp_disc}
\vspace{-0.5em}

Discriminative power indicates how well a metric can tell systems apart. Zhou et al. \cite{zhou2013reliability} measure discriminative power by counting the number of significantly different system pairs. We apply this same method to measure the discriminative power of our proposed metrics, using Tukey's HSD test as the statistical hypothesis test. 
\textbf{Table \ref{Tab:Discriminative}} shows the  number of significantly different pairs for baseline and our proposed metrics when we use
95\% and 99\% significance thresholds. 

We observe that our metrics have higher discriminative power than baselines. Increasing $\alpha$ tends to increase discriminative power across test collections. 

\begin{table}[!htb]
\centering
\setlength\tabcolsep{2.4pt}
  \begin{tabular}{ |l|l|c|c|c|c|c|c|c|c|}
     \hline
     & & \multicolumn{2}{c|}{\bf TREC-5} & \multicolumn{2}{c|}{\bf TREC-8} & \multicolumn{2}{c|}{\bf WT14} & \multicolumn{2}{c|}{\bf DLT20}\\ \cline{3-10}
    {\bf Metric} & {\bf $\alpha$} & 95\% & 99\% & 95\% & 99\% & 95\% & 99\% & 95\% & 99\% \\
     \hline
    P@100 & 0.0 & 598 & 406 & 3666 & 2973 & 218 & 174 & 61 & 15  \\
    $P@100_{Rareness}$ & 0.5 & 680 & 457 & 3778 & 3050 & 213 & 168 & 173 & 24\\
    $P@100_{Rareness}$ & 1.0 & 728 & 476 & 3825 & 3079 & 213 & 168 & 237 & 82  \\
    \hline
    AP & 0.0 & 541 & 334 & 3731 & 2976 & 211 & 169 & 320 & 169  \\
    $AP_{Rareness}$ & 0.5 & 632 & 404 & 3915 & 3209 & 216 & 168 & 352 & 209\\
    $AP_{Rareness}$ & 1.0 & 701 & 467 & 4048 & 3363 & 214 & 170 & 376 & 238  \\
    
     \hline
  \end{tabular}
\caption{Discriminative power of metrics for 95\% and 99\% significance thresholds.  The highest score for each collection and significance threshold is written in \textbf{bold}. Note that the total number of system pairs are 1830, 8256, 406, 2016 for TREC-5, TREC-8, WT14, DTL20, respectively. }
  \label{Tab:Discriminative}
\end{table}


\subsection{Stability}\label{sec_exp_rel}
\vspace{-0.5em}




If an evaluation methodology is reliable, the measured performance of systems should be stable, i.e., should not change dramatically under different conditions. In order to measure the stability of metrics, we adopt Buckley and Voorhees \cite{buckley2017evaluating}'s approach. We first sample $T$ topics and calculate system scores on the sampled topic set only. Next, we compare each pair of systems to see which performs better. After repeating this process $R$ times, we assess the stability of the comparison over the $R$ trials. For example, imagine one system outperforms another in 700/1000 trials, yielding a stability score of 0.7 for that pair.  
We take the average stability scores of all pairs as the overall metric stability. In our experiments, we arbitrarily set $T$ to the half of the topic set in each collection (i.e., 22 ($=\lfloor45/2\rfloor$) for DLT20 and 25 ($=50/2$) for the others). We set the number of trials $R=1000$ but observed that the results largely converged after 100 trials. Results for baselines vs.\ proposed metrics are shown in \textbf{Table \ref{Tab:reliability}}.  $P@100_{Rareness}$ and $AP_{Rareness}$ yield a higher stability score in all cases vs.\ their classic counterparts. 

\begin{table}[ht]
\centering
  \begin{tabular}{ |l|c|c|c|c|c|  }
     \hline
     {\bf Metric} & {\bf $\alpha$} & {\bf TREC-5} & {\bf TREC-8} & {\bf WT14} & {\bf DLT20}\\
     \hline
    P@100 & 0.0 & 0.532 & 0.545 & 0.635 & 0.071 \\
    $P@100_{Rareness}$ & 0.5 & 0.642 & 0.628 & 0.716 & 0.128 \\
    $P@100_{Rareness}$ & 1.0 & 0.709 & 0.684 & 0.768 & 0.179 \\ 
    \hline
    AP & 0.0 & 0.513 & 0.580 & 0.433 & 0.425 \\
    $AP_{Rareness}$ & 0.5 & 0.578 & 0.623 & 0.517 & 0.444 \\
    $AP_{Rareness}$ & 1.0 & 0.633 & 0.656 & 0.585 & 0.466 \\
     \hline
  \end{tabular}
  \caption{Metric stability scores. Our metrics are most stable across collections.} 
  \label{Tab:reliability}
\vspace{-2em}
\end{table}

\subsection{Impact of Number of Systems}\label{sec_exp_system}
\vspace{-0.5em}

As retrieval-rarity of documents depends on the participating systems,  
system scores and rankings might change when we use a different set of systems to calculate the rarity scores of documents. 
To test how scores of systems change as the systems to be evaluated vary, we conduct the experiment described in \textbf{Algorithm \ref{alg_exp}}. In particular, we first rank all systems [Line 1]. Then we randomly pick $N$ number of systems [Line 5] and rank them [Line 6]. Subsequently, we get how these $N$ systems are ranked initially (i.e., when all systems are used) [Line 7] and calculate the $\tau$ score between these two rankings [Line 8]. We repeat this process 1000 times [Lines 3-9] and calculate the average $\tau$ score [Line 10]. \textbf{Table \ref{Tab:exp_system_number}} shows the results for  $N = 2^j, j \in [1-6]$ in TREC-8. We observe that correlation scores are generally very high, suggesting that rankings of systems are stable even though we use different sets of systems.

\begin{algorithm}
\small
\caption{Experiment to Analyze Impact of  Using $N$ Participants}\label{alg_exp}
\ \ \ \ \ \textbf{Input:} $P \gets$ The whole participant list \\
\hspace*{\algorithmicindent} \ \ \ \ \ \ \ \ \  $N \gets$ The number of selected systems
\begin{algorithmic}[1]
\State {$R_o \gets$ rank systems in $P$}
\State {$\tau_N \gets$ 0}
\State $trials \gets 1000$
\ForAll {$trials$}
\State {$p_N \gets$ randomly sample N systems from P}
\State {$r_N \gets$ rank systems in $p_N$}
\State {$R_N \gets$ filter systems $\in p_N$ from $R_o$}
\State {$\tau_N \gets \tau_N + \tau\_correlation(R_N,r_N)$}
\EndFor 
\State {$\tau_N \gets \tau_N$ / $trials$}
\end{algorithmic}
\end{algorithm}



\begin{table}[!htb]
  \vspace{-2em}
\centering
  \begin{tabular}{ |l||c|c|c|c|c|c|  }
     \hline
     \ Metrics & N=2 & N=4 & N=8 & N=16 & N=32 & N=64 \\
     \hline
     
    \ $P@100_{Rareness}\ \  (\alpha =1)$ & \ 0.970\ \  & \ 0.976\ \  & \ 0.983\ \  & \ 0.987\ \  & \ 0.992\ \  & \ 0.995\ \  \\
    \ $AP_{Rareness} (\alpha =1)$ & \ 0.970\ \  & \ 0.984\ \  & \ 0.985\ \  & \ 0.989\ \  & \ 0.993\ \  & \ 0.996\ \  \\
     \hline
  \end{tabular}
  \caption{Impact of number of systems   based on the experimental setup explained in Algorithm \ref{alg_exp}. We use TREC-8 for this experiment.} 
  \label{Tab:exp_system_number}
  \vspace{-4em}
\end{table}

\subsection{Qualitative Analysis}\label{sec_exp_qual}
\vspace{-0.5em}


To better understand the nature of rarely-retrieved documents, we conducted the following qualitative analysis. We randomly selected six TREC-8 topics, computed the rarity $R(d)$ of each relevant document $d$, and then selected five documents with varying rarity scores. 
We manually analyzed how document relevance changes depending on rarity. In general, while commonly retrieved documents appear focused on the search topic, rarely retrieved documents differ in focus but still contain relevant passages.

\textbf{Table \ref{Tab:qual}} presents manually analyzed documents for topic 431, whose narrative states the information need: ``latest developments in robotic technology''. The relevant document FBIS4-44815 with minimal rarity is entitled, ``Germany: Automation, Robotics Seen as Keys in Industrial'', which seems directly relevant to the information need. In contrast, relevant document FBIS3-38782 ($R(d)=0.81$) only indirectly mentions that a robot can be used for underwater photography, with the title ``BND Warns Against Nuclear Terrorists''.

If rarely retrieved documents are less relevant, why reward their retrieval?  First, while the observation above may hold when all systems are roughly comparable, this is not always true. For example, manual runs have long been advocated in evaluation campaigns because they tend to differ markedly from automated runs and find relevant documents that other systems miss. In general, we cannot tell whether outlier systems are brilliant or remedial without human labels \cite{soboroff2001ranking}. Second, our goal in this work is to encourage systems that diverge from the pack, with the hope that such divergence will correspond to improvement. The nature of research is that some amount of failure often precedes success, and that making larger departures is important to create a potential for larger improvements. Third, even if we assume a user-centered view, finding additional, less relevant documents can still be important in various cases: when there are few relevant documents, in a ``total recall'' task setting \cite{roegiest2015trec} or pooling \cite{spark1975report}, or as input to a rank fusion ensemble model \cite{nuray2006automatic}. We discuss these further in the next section.

\begin{table}[!htb]
\small
\centering
  \begin{tabular}{ |p{1.8cm}|l|p{8.4cm}| }
    \hline
    Document ID & Rareness & Relevant Content\\
    \hline
    FBIS3-38782 & 0.81 & One of the trapeze-like wings had broken off, the nose was missing, and in the dull gray water of Lake Constance even a diving robot of the ``Sear Rover'' type could send only diffuse video pictures from 159 meters below the surface of the lake.\\
    \hline
    LA020889-0003 & 0.62 & A Japanese robot named Wabot II tickles a keyboard to produce original music as part of an exhibit at the Chicago Museum of Science and Industry.\\
    \hline
    LA102589-0109 & 0.41 & The trucks, equipped with robotic arms that hoist and empty containers, will collect trash every week and recyclable items twice monthly.\\
    \hline
    LA092189-0061 & 0.22 & Industrially they use robots for welding, painting or picking and placing items, for example.\\
    \hline
    FBIS4-44815 & 0.08 & New applications for service robots are opening up also in medicine and rehabilitation, in care for the aged and handicapped, in bureaus and logistics, in municipal activities, in households, in hobbies and recreation.\\
    \hline
  \end{tabular}
  \caption{Analyzed documents for topic 431. The relevant content column corresponds to sentences that might fulfill the information need. If there are multiple useful sentences in a document, the most informative one is selected.} 
  \label{Tab:qual}
\vspace{-2em}
\end{table}

\section{Discussion and Limitations} \label{sec_disc} \label{sec_limit}
\vspace{-1em}

In this section, we discuss various aspects and limitations of our work: motivation and concept of ``innovation’’ metrics (Section \ref{sec_disc_conc}), proposed methods (Section \ref{sec_prop}), our experimental design and findings (Section \ref{sec_disc_exp}), and potential impacts and directions for future work (Section \ref{sec_disc_expec}).

\subsection{Concept and Motivation} \label{sec_disc_conc}
\vspace{-0.5em}

We envision potential benefit from stimulating greater diversity in document ranking methods. In regard to evaluation campaigns, we suggest the field would benefit if participants built upon a wider range of existing methods and/or investigated more radical departures from those methods. While today's evaluation campaigns are already healthy and vibrant, we believe it could be fruitful: 1) to reflect on, assess, and discuss as a community the ways in which we might further strengthen evaluation campaigns; 2)  to focus on the diversity of approaches and innovation in particular, and how to promote higher-risk research with potential for greater gains; and 3) to operationalize metrics by which we might measure and optimize for such innovation in evaluation campaigns. Potential counter-arguments could be that: a) campaign steering committees are already doing (1) and don’t need larger community engagement in it; b) innovation is a complex construct that is best left to organic processes rather than trying to ``force’’ it through explicit optimization; and c) one can argue that research construed as incremental is actually instrumental  (i.e., small steps and minor variants can add up over time to large advances). Such discussion and debate seem healthy for a community, regardless of the outcome. 

One controversial aspect of our work is the proposal of IR evaluation metrics that explicitly seek to optimize something other than retrieval quality for the user. In particular, the metrics we propose reward systems for retrieving relevant documents missed by other systems, but there is no obvious reason a user would prefer such rare relevant documents over common ones.  In fact, less retrieved documents may tend to be less relevant on average and thus aptly lower-ranked (Section \ref{sec_exp_qual}).
In fact, prior work in meta-ranking (aka rank fusion) has exploited the number of systems that retrieve a given document as a useful feature in estimating document relevance \cite{nuray2006automatic}. However, our goal of promoting greater community diversity of ranking methods is not a user-oriented metric, but a field-oriented metric. Moreover, in seeking to promote higher-risk research, we may need to explore a variety of methods yielding sub-par results for the user before we discover a novel method that does provide a transformative advance. For example, years of research on (then) sub-par neural networks was necessary before yielding today’s state-of-the-art deep learning methods \cite{onal2018neural}. 

Rewarding retrieval of rare relevant documents also has the potential to improve meta-ranking (aka rank fusion or ensemble ranking) and pooling \cite{spark1975report}. For instance, ensemble models benefit from a diverse set of input systems that complement each other's shortcomings. Thus, including input systems that find unique relevant documents could boost ensemble performance. Pooling similarly benefits from the diversity of participating systems so that the pool finds as many relevant documents as possible. This helps to ensure that the pool is reusable for future systems using innovative approaches. Our metrics could thus encourage more diverse systems to improve meta-ranking and pooling. In the other direction, recall measures for those tasks might also be repurposed to measure and promote overall diversity and innovation of ranking approaches. 

\subsection{Proposed Metrics} \label{sec_prop}
\vspace{-0.5em}

The two specific innovation metrics we propose have a variety of limitations and represent only the tip of the iceberg of better innovation metrics. We expect future work will propose better metrics that surpass ours.

As noted above, the notion of innovation is a complex construct. Our metrics that reward retrieval of relevant documents missed by other systems are clearly crude metrics for quantifying such a complex construct. To the best of our knowledge, ours is the first metric for measuring and promoting such innovation, but the first effort seldom represents the only or best way. More sophisticated future work by others could model this construct with greater detail and fidelity. 

While we have suggested combining $Rel(d_i)$ and $Rel(d) \cdot R(d)$ together into a single mixture for simplicity, an evaluation campaign could also use these as separate and complementary official metrics, akin to evaluating precision vs.\ recall separately rather than fusing them together into a single f-measure metric. On the other hand, our mixture approach can also be seen as an easy way to generalize existing metrics to consider additional aspects of performance. Because our modified metrics revert to their standard counterpart metrics when $\alpha=0$, generalization allows use in that original, more restricted setting while also permitting greater flexibility in incorporating additional factors when $\alpha>0$. While we focus on generalizing existing metrics to include consideration of document rarity, other researchers might incorporate other aspects of system performance into traditional metrics using similar linear mixtures.   

At a more mundane level, because our metrics are bounded by $[0,2)$, it may be useful to renormalize them to a more standard $[0,1]$ range.  While this might be done to values post hoc, hindsight instead suggests two minor revisions to formulas for future use.  First, re-define rarity as $R’(d) = 1 - \frac{S_d - 1}{S - 1} \in [0,1]$ for $S, S_d >= 1$, maximized when $S_d = S$. Second, re-define $P’@K_{rareness}$ as: 
\vspace{-0.5em}
\begin{equation}
P’@K_{rareness} = \frac{1}{k}\sum_{i=1}^{k} \ \Bigl[ (1 - \alpha) Rel(d_i) + \alpha \, Rel(d_i) \, R(d_i) \Bigr] 
\end{equation}
where we now constrain $\alpha \in [0,1]$ as a probability. This mixture model formulation directly bounds $P’@K_{rareness} \in [0,1]$. 

Our metrics assume linearity in: 1) how we quantify rarity $R(d)$; and 2) the mixture model between the classic metric and rarity. If we consider IR's rich history exploring many variant functions for inverse-document frequency (IDF) to weight rare terms \cite{salton1988term}, one could imagine similarly exploring many other weighting functions for rarity. Regarding the mixture model, while we have assumed a fixed $\alpha$ across topics, future work might also investigate a hyperparameter approach (akin to Dirichlet smoothing \cite{zhai2017study}) to intelligently vary $\alpha$ per topic in relation to per topic factors, such as the number of relevant documents.

Yet another idea would be to incorporate document importance alongside rarity in the reward metric for innovation. Intuitively, finding a relevant document that other systems miss is more important when there are few relevant documents in total. As an example, assume for some topic that a given relevant document is only retrieved by a single system. If there are only two relevant documents in total, finding that second relevant document may be vital to satisfying a user's information need. On the other hand, if there were 100 relevant documents, finding the $100^{th}$  document may provide minimal further value. This would suggest extending the metric to consider the number of relevant documents for each topic. 

Finally, our use of P@K and AP assumes binary relevance judgments.  Future work could extend innovation metrics to graded relevance judgments.

\subsection{Experimental Design and Findings} \label{sec_disc_exp}
\vspace{-0.5em}

While we evaluated over four test collections to assess generality, we did not explore the properties of these test collections in detail, or how those varying properties could impact our findings.  In addition, expanding our coverage to further test collections could further assess the robustness of findings.  Finally, it could be useful to conduct a qualitative inspection of the meta-data descriptions of the best-performing systems (submitted by participants along with their TREC runs) in order to assess the correlation between system descriptions vs.\ which systems perform best when scored by our innovation metrics.




\subsection{Expected Use and Impact} \label{sec_disc_expec}
\vspace{-0.5em}

Imagine our metrics were adopted by an evaluation campaign and one or more participating systems sought to optimize them.  Beyond the broad goal of promoting higher-risk research and accelerating field innovation, this would be expected to specifically lead to more diverse document rankings. Assuming a fixed evaluation budget (i.e., the number of documents that human judges will review), less overlap across document rankings would mean that we could only pool to a lower depth for the same cost. However, whether this would lead to a more or less complete document pool remains an open, empirical question, likely dependent on the setting of $\alpha$ used. For evaluation campaigns that permit participants to submit multiple runs and distinguish an ``official’’ run (contributing to pooling) vs.\ additional runs (scored by the official run pool), whether official vs.\ additional runs would be used to set $S_d$ would also impact subsequent findings.   

A well-known issue in IR is the reusability of pools. A very different system might find relevant documents all other systems missed, but if it did not participate in the pool, it would be penalized in evaluation rather than rewarded. Similarly, when we quantify rarity $R(d)$ based on participating systems, there are questions of reusability for future systems evaluating on an existing pool. Moreover, we would expect that a system optimizing for such rarity would be even more likely to run into this problem in practice. Another common distinction made is between methods to create reusable test collections (e.g., pooling) vs.\ methods to efficiently rank a current set of systems (e.g., StatAP \cite{pavlu2007practical} and MTC \cite{carterette2006minimal}). Similarly, our rarity metrics will return different scores depending on the other participating systems in the pool. A limitation of our work is that we only rank systems participating in a shared-task, leaving  study of reusability for future work.

\section{Related Work}\label{sec_rel}
\vspace{-1em}


To the best of our knowledge, no existing IR evaluation metrics consider the innovativeness of systems.  
%
%
While we frame this {\em wrt}.\ rarely-retrieved documents, 
prior work has usefully designed metrics to evaluate systems reliably with missing judgments, such as Bpref \cite{bpref} and infAP \cite{infap}. 
These metrics aim to predict the performance of systems with incomplete judgments. In contrast, our focus is to promote innovation in document ranking methods.

To handle missing judgments, a number of studies have explored how to select documents to be judged such as Move-To-Front \cite{cormack1998efficient} and MaxMean \cite{losada2017multi}. These studies aim to maximize the number of relevant documents because unjudged documents are assumed to be non-relevant. As a document is more likely to be relevant if retrieved by many systems, commonly-retrieved documents are more likely to be judged than rarely retrieved ones.  
However, in contrast to these document selection methods, we assign more weight to rarely-retrieved ones. 


In modifying P@K and AP, we have followed standard practice in aggregating scores over topics using a simple arithmetic average. However, various other aggregate statistics have been proposed.  
Robertson \cite{robertson2006gmap} asserts that the impact of hard topic scores is diminished on the overall score with the arithmetic mean. He thus recommends geometric mean instead. 
Ravana and Moffat \cite{gmap}  show that 
 geometric mean average precision (GMAP) is
better at handling variability in topic difficulty than arithmetic mean average precision (MAP).
Mizzaro \cite{mizzaro2008} proposes normalized mean average precision (NMAP), which takes into account topic difficulty. He defines topic difficulty as 1-(average AP score). 
Unlike these studies, we focus on retrieval difficulty at the document level. In addition, prior studies on topic difficulty work on how to aggregate traditional IR metrics. 

As noted earlier, while we assumed binary relevance judgments and modify only P@K and AP metrics, many other metrics exist, beyond binary relevance, that could be extended to innovation. Prominent examples include normalized discounted cumulative gain (nDCG) \cite{nDCG}, and rank biased precision (RBP) \cite{rbp}, which assume that users will examine documents in the retrieval order and might stop examining whenever their information need is satisfied. Such rank-based metrics ascribe more weight to documents at higher ranks. Other important evaluation metrics include miss (i.e., the fraction of non-retrieved documents that are relevant) \cite{heine1984information}, fallout \cite{kraft1978evaluation}, expected reciprocal rank \cite{chapelle2009expected}, weighted reciprocal rank \cite{eguchi2002overview}, and O-measure \cite{sakai2006task}.


Prior work has also proposed metrics rewarding the diversity within a single document ranking in relation to novelty and coverage of different topic facets. 
For instance, Zhai et al. \cite{zhai2015beyond} propose three metrics -- subtopic recall metric (S-recall), subtopic precision (S-precision) and weighted subtopic precision (WS-precision) -- that consider redundancy in ranked lists. Clarke et al. \cite{clarke2008novelty} extend nDCG by rewarding novelty and covering multiple topic aspects. In contrast, we quantify diversity across systems rather than within a single ranked list. In particular,  we reward systems for retrieving relevant documents that other systems miss. 


\section{Conclusion}\label{sec_conc}
\vspace{-1em}

We propose a new class of IR evaluation metrics designed to promote exploration of higher-risk, more radical departures from current state-of-the-art methods. These  ``innovation metrics'' reward retrieval of relevant documents missed by other systems.  The key intuition is that finding relevant documents missed by other systems suggests a markedly different approach. More specifically, we generalize classic Precision@K and Average Precision metrics via a simple mixture-weight parameter controlling the relative reward for finding relevant documents other systems miss. Setting this to zero reverts to the original metric. 

Experiments over four TREC collections show that our proposed metrics yield different system rankings compared to the existing metrics. In particular, we observe a steady decrease in rank correlation with official system rankings
as reward increases for finding rare, relevant documents.
These results are further supported by a controlled, synthetic data experiment, as well as qualitative analysis. Collectively, results suggest that if our metrics were adopted in practice, participants would be incentivized to retrieve more diverse relevant documents, with the potential to spur further innovation in the field. Finally, we also show that our metrics provide higher discriminative power and evaluation stability than the standard Precision@K and Average Precision metrics that we modify.

To the best of our knowledge, ours is the first proposal of IR evaluation metrics designed to explicitly measure and promote innovation in ranking methods. That said, the first attempt at any endeavor is seldom the only or best way to accomplish it. Our two proposed metrics have a variety of limitations and represent only the tip of the iceberg for imagining this new class of innovation metrics. Consequently, we expect future metrics will be proposed that surpass ours in better modeling the complex construct of innovation, and in doing so, will further advance the cause of promoting innovation in ranking methods. 

~\\\noindent 
{\small {\bf Acknowledgments}. 
We thank the reviewers for their valuable feedback. This research was supported in part by the Scientific and Technological Research Council of Turkey (TUBITAK) ARDEB 3501 (Grant No 120E514) and by Good Systems\footnote{\url{https://goodsystems.utexas.edu}}, a UT Austin Grand Challenge to develop responsible AI technologies. Our opinions are our own.
}

\begin{thebibliography}{10}
\providecommand{\url}[1]{\texttt{#1}}
\providecommand{\urlprefix}{URL }
\providecommand{\doi}[1]{https://doi.org/#1}

\bibitem{brown2020language}
Brown, T., Mann, B., Ryder, N., Subbiah, M., Kaplan, J.D., Dhariwal, P.,
  Neelakantan, A., Shyam, P., Sastry, G., Askell, A., et~al.: Language models
  are few-shot learners. Advances in neural information processing systems
  \textbf{33},  1877--1901 (2020)

\bibitem{bpref}
Buckley, C., Voorhees, E.M.: Retrieval evaluation with incomplete information.
  In: Proceedings of the 27th annual international ACM SIGIR conference on
  Research and development in information retrieval. pp. 25--32 (2004)

\bibitem{buckley2017evaluating}
Buckley, C., Voorhees, E.M.: Evaluating evaluation measure stability. In: ACM
  SIGIR Forum. vol.~51, pp. 235--242. ACM New York, NY, USA (2017)

\bibitem{carterette2006minimal}
Carterette, B., Allan, J., Sitaraman, R.: Minimal test collections for
  retrieval evaluation. In: Proceedings of the 29th annual international ACM
  SIGIR conference on Research and development in information retrieval. pp.
  268--275 (2006)

\bibitem{chapelle2009expected}
Chapelle, O., Metlzer, D., Zhang, Y., Grinspan, P.: Expected reciprocal rank
  for graded relevance. In: Proceedings of the 18th ACM conference on
  Information and knowledge management. pp. 621--630 (2009)

\bibitem{clarke2008novelty}
Clarke, C.L., Kolla, M., Cormack, G.V., Vechtomova, O., Ashkan, A.,
  B{\"u}ttcher, S., MacKinnon, I.: Novelty and diversity in information
  retrieval evaluation. In: Proceedings of the 31st annual international ACM
  SIGIR conference on Research and development in information retrieval. pp.
  659--666 (2008)

\bibitem{wt14}
Collins-Thompson, K., Macdonald, C., Bennett, P., Diaz, F., Voorhees, E.M.:
  Trec 2014 web track overview. Tech. rep., MICHIGAN UNIV ANN ARBOR (2015)

\bibitem{cormack1998efficient}
Cormack, G.V., Palmer, C.R., Clarke, C.L.: Efficient construction of large test
  collections. In: Proceedings of the 21st annual international ACM SIGIR
  conference on Research and development in information retrieval. pp. 282--289
  (1998)

\bibitem{craswell2021overview}
Craswell, N., Mitra, B., Yilmaz, E., Campos, D.: Overview of the trec 2020 deep
  learning track. arXiv preprint arXiv:2102.07662  (2021)

\bibitem{eguchi2002overview}
Eguchi, K., Oyama, K., Ishida, E., Kando, N., Kuriyama, K.: Overview of the web
  retrieval task at the third ntcir workshop. In: NTCIR. Citeseer (2002)

\bibitem{harman1996overview}
Harman, D., Voorhees, E.: Overview of the fifth text retrieval conference
  (trec-5). In: Information Technology: The Fifth Text REtrieval Conference
  (TREC-5), D. Harman and E. Voorhees, eds., National Institute of Standards
  and Technology Special Publication. pp. 500--238 (1996)

\bibitem{trec8}
Hawking, D., Voorhees, E., Craswell, N., Bailey, P., et~al.: Overview of the
  trec-8 web track. In: TREC (1999)

\bibitem{heine1984information}
Heine, M.: Information-retrieval from classical databases from a
  signal-detection standpoint-a review. Information Technology-Research
  Development Applications  \textbf{3}(2),  95--112 (1984)

\bibitem{nDCG}
J{\"a}rvelin, K., Kek{\"a}l{\"a}inen, J.: Ir evaluation methods for retrieving
  highly relevant documents. In: ACM SIGIR Forum. vol.~51, pp. 243--250. ACM
  New York, NY, USA (2017)

\bibitem{kraft1978evaluation}
Kraft, D.H., Bookstein, A.: Evaluation of information retrieval systems: A
  decision theory approach. Journal of the American Society for Information
  Science  \textbf{29}(1),  31--40 (1978)

\bibitem{losada2017multi}
Losada, D.E., Parapar, J., Barreiro, A.: Multi-armed bandits for adjudicating
  documents in pooling-based evaluation of information retrieval systems.
  Information Processing \& Management  \textbf{53}(5),  1005--1025 (2017)

\bibitem{mizzaro2008}
Mizzaro, S.: The good, the bad, the difficult, and the easy: something wrong
  with information retrieval evaluation? In: European Conference on Information
  Retrieval. pp. 642--646. Springer (2008)

\bibitem{rbp}
Moffat, A., Zobel, J.: Rank-biased precision for measurement of retrieval
  effectiveness. ACM Transactions on Information Systems (TOIS)
  \textbf{27}(1),  1--27 (2008)

\bibitem{nuray2006automatic}
Nuray, R., Can, F.: Automatic ranking of information retrieval systems using
  data fusion. Information processing \& management  \textbf{42}(3),  595--614
  (2006)

\bibitem{onal2018neural}
Onal, K.D., Zhang, Y., Altingovde, I.S., Rahman, M.M., Karagoz, P., Braylan,
  A., Dang, B., Chang, H.L., Kim, H., McNamara, Q., et~al.: Neural information
  retrieval: At the end of the early years. Information Retrieval Journal
  \textbf{21}(2),  111--182 (2018)

\bibitem{pavlu2007practical}
Pavlu, V., Aslam, J.: A practical sampling strategy for efficient retrieval
  evaluation. College of Computer and Information Science, Northeastern
  University  (2007)

\bibitem{gmap}
Ravana, S.D., Moffat, A.: Exploring evaluation metrics: Gmap versus map. In:
  Proceedings of the 31st annual international ACM SIGIR conference on Research
  and development in information retrieval. pp. 687--688 (2008)

\bibitem{robertson2006gmap}
Robertson, S.: On gmap: and other transformations. In: Proceedings of the 15th
  ACM international conference on Information and knowledge management. pp.
  78--83 (2006)

\bibitem{roegiest2015trec}
Roegiest, A., Cormack, G.V., Clarke, C.L., Grossman, M.R.: Trec 2015 total
  recall track overview. In: TREC (2015)

\bibitem{sakai2006task}
Sakai, T.: On the task of finding one highly relevant document with high
  precision. Information and Media Technologies  \textbf{1}(2),  1025--1039
  (2006)

\bibitem{salton1988term}
Salton, G., Buckley, C.: Term-weighting approaches in automatic text retrieval.
  Information processing \& management  \textbf{24}(5),  513--523 (1988)

\bibitem{soboroff2001ranking}
Soboroff, I., Nicholas, C., Cahan, P.: Ranking retrieval systems without
  relevance judgments. In: Proceedings of the 24th annual international ACM
  SIGIR conference on Research and development in information retrieval. pp.
  66--73 (2001)

\bibitem{spark1975report}
Spark-Jones, K.: Report on the need for and provision of an'ideal'information
  retrieval test collection. Computer Laboratory  (1975)

\bibitem{voorhees2000variations}
Voorhees, E.M.: Variations in relevance judgments and the measurement of
  retrieval effectiveness. Information processing \& management
  \textbf{36}(5),  697--716 (2000)

\bibitem{infap}
Yilmaz, E., Aslam, J.A.: Estimating average precision with incomplete and
  imperfect judgments. In: Proceedings of the 15th ACM international conference
  on Information and knowledge management. pp. 102--111 (2006)

\bibitem{tauap}
Yilmaz, E., Aslam, J.A., Robertson, S.: A new rank correlation coefficient for
  information retrieval. In: Proceedings of the 31st annual international ACM
  SIGIR conference on Research and development in information retrieval. pp.
  587--594 (2008)

\bibitem{zhai2015beyond}
Zhai, C., Cohen, W.W., Lafferty, J.: Beyond independent relevance: methods and
  evaluation metrics for subtopic retrieval. In: ACM SIGIR forum. vol.~49,
  pp.~2--9. ACM New York, NY, USA (2015)

\bibitem{zhai2017study}
Zhai, C., Lafferty, J.: A study of smoothing methods for language models
  applied to ad hoc information retrieval. In: ACM SIGIR Forum. vol.~51, pp.
  268--276. ACM New York, NY, USA (2017)

\bibitem{zhou2013reliability}
Zhou, K., Lalmas, M., Sakai, T., Cummins, R., Jose, J.M.: On the reliability
  and intuitiveness of aggregated search metrics. In: Proceedings of the 22nd
  ACM international conference on Information \& Knowledge Management. pp.
  689--698 (2013)

\end{thebibliography}

\end{document}